\documentclass[12pt, reqno]{amsart}
\usepackage{amsmath, amsthm, amscd, amsfonts, amssymb, graphicx, color}
\usepackage[bookmarksnumbered, colorlinks, plainpages]{hyperref}
\hypersetup{colorlinks=true,linkcolor=red, anchorcolor=green, citecolor=cyan, urlcolor=red, filecolor=magenta, pdftoolbar=true}

\theoremstyle{definition}

\theoremstyle{remark}

\numberwithin{equation}{section}

\begin{document}

\setcounter{page}{1}

\title[   Logical Phenomena and the Cognitive System ]{ The Interplay Between Logical Phenomena and the Cognitive System of the Mind }
\author[Kazem Haghnejae Azar]{Kazem Haghnejae Azar}

\address{Mathematics Department, Faculty of Science, University of Mohaghegh Ardabili, Ardabil, Iran}
\email{\textcolor[rgb]{0.00,0.00,0.84}{haghnejad@uma.ac.ir}}


\begin{abstract}
In this article, we employ mathematical concepts as a tool to examine the phenomenon of consciousness experience and logical  phenomena. Through our investigation, we aim to demonstrate that our experiences, while not confined to limitations, cannot be neatly encapsulated within a singular collection. 
Our conscious experience emerges as a result of the developmental and augmentative trajectory of our cognitive system. As our cognitive abilities undergo refinement and advancement, our capacity for logical thinking likewise evolves, thereby manifesting a heightened level of conscious experience. The primary objective of this article is to embark upon a profound exploration of the concept of logical experience, delving into the intricate process by which these experiences are derived from our mind.

\end{abstract} \maketitle

\contentsline {section}{\numberline {1}Preliminaries and Introduction}{}{section.1}
\contentsline {section}{\numberline {2}Exploring the Root of Conscious Experience}{}{section.2}
\contentsline {section}{\numberline {3} Mathematical Approaches in the Study of Consciousness Experience}{}{section.3}
\contentsline {section}{\numberline {4}Extracting Logical and Conceptual Phenomena from Experiential Realms}{}{section.4}
\contentsline {section}{\numberline {5}Corollary}{}{section.5}
\contentsline {section}{\numberline {6}References}{}{section.6}

\section{Preliminaries and Introduction }

To gain a comprehensive understanding of the intricate relationship between the chemical and physical effects of the brain and the experience of consciousness, it is indeed crucial to delve into the process of transforming a non-living chemical substance into a living entity and the subsequent creation of a cognitive system, see Azar (2023)\cite{Azar}.
The transition from non-living matter to a living organism involves a complex series of events, including chemical reactions, self-organization, and the emergence of biological structures. These processes give rise to the fundamental building blocks of life, such as proteins, nucleic acids, and membranes, which form the basis for the intricate machinery of a living system.
Within this living entity, the creation of a cognitive system involves the organization and integration of various components, including neural networks, sensory organs, and information processing mechanisms. The precise mechanisms through which consciousness arises from these physical and chemical processes remain a subject of ongoing scientific inquiry and philosophical contemplation.
By exploring and unraveling the process of converting non-living chemical substances into living organisms and the subsequent emergence of cognitive systems, we can gain valuable insights into the nature of consciousness and the intricate interplay between the physical, chemical, and cognitive aspects of our existence. This exploration offers a pathway towards a deeper understanding of the profound phenomena underlying consciousness and the complex dynamics that give rise to our subjective experiences.
As expounded by Pross (2021)\cite{23pp}, the attainment of a sentient chemical system possessing the capacity to evolve and adapt in order to optimize its interaction with the environment assumes paramount importance for the augmentation of persistence. This protracted process, unfolding over a vast expanse of time, ultimately engendered the advent of the bacterial cell, signifying a momentous milestone in the evolutionary trajectory towards more enduring manifestations.
According to Pascal and  Pross (2023)\cite{Pascal and  Pross}, even bacteria, which represent the simplest form of prokaryotic life, possess the ability to sense and respond to a wide range of environmental signals. These organisms can sense their internal conditions, coordinate with neighboring organisms, and activate complex response systems to address ongoing challenges, see Shapiro (2007)\cite{24shap1}.

Cognitive systems emerge through the interaction of chemical sensors within certain compounds with the external environment and internal processes. As this process continues, it can give rise to unique characteristics in certain chemical compounds, resulting in the formation of living organisms.
The ongoing development of this process can lead to the emergence of organisms with more advanced cognitive mechanisms, which we commonly refer to as the mind. The mind encompasses a range of cognitive abilities such as perception, memory, learning, and decision-making. These cognitive capabilities are closely linked to the biological structure of organisms, including the development of complex nervous systems and brain structures that support higher-order cognitive functions.
It is through the interplay of these cognitive mechanisms and the biological structures that organisms are capable of exhibiting more sophisticated behaviors, adapting to their environment, and engaging in complex cognitive processes.

Indeed, the experience of consciousness can be seen as an evolutionary development of the cognitive mechanism in living organisms. This process began with simple living systems and may have even emerged from non-living chemical systems. Over time, this cognitive mechanism has evolved and become more complex, culminating in the intricate consciousness experienced by humans and other advanced organisms.
Consciousness is believed to have emerged gradually in the course of biological evolution. It likely began in simple living systems with rudimentary forms of perception and awareness. Even single-celled organisms exhibit basic sensory responses and the ability to navigate their environment, suggesting the presence of some level of consciousness.
As life forms evolved and became more complex, so did their cognitive abilities. The development of nervous systems and the emergence of more sophisticated sensory organs allowed for enhanced information processing and perception. This evolutionary trajectory led to the gradual refinement and expansion of conscious experiences.

The human brain comprises approximately one hundred billion neurons, each establishing connections with tens of thousands of other neurons. These connections occur through electrochemical junctions known as synapses, enabling communication between neurons. Through this intricate network, neurons collaborate and form interactive networks that collectively regulate various functions of the human body, ranging from fundamental processes like breathing to complex activities such as composing music.
At any given moment, the brain processes a vast array of sensory information and integrates it with stored information from memory. This amalgamation of inputs allows the brain to generate perceptions and thoughts. It is through this dynamic interplay between sensory input and memory that our conscious experiences and cognitive processes emerge. The brain's ability to combine and analyze information in real-time contributes to our capacity for perception, decision-making, and the construction of coherent thoughts.

In essence, the brain perpetually processes an immense influx of sensory information from the external realm. It amalgamates this incoming sensory data with stored information retrieved from memory, giving rise to perceptions and thoughts. This integration of sensory input and memory retrieval contributes to our overarching cognitive experience and the formation of conscious awareness.
As elucidated by Allen (2009)\cite{1aa}, the brain emerges as a product shaped by a multitude of contextual factors, encompassing the phylogenetic, somatic, genetic, ecological, demographic, and ultimately, culturo-linguistic dimensions within which it evolved.

The mathematical concepts employed in this article are elementary and accessible to individuals with a basic understanding of mathematics.

\section{Exploring the Root of Conscious Experience}

Pross and Pascal (2017) \cite{23p}  expound upon a compelling argument, establishing that chemical systems on the trajectory towards heightened dynamic kinetic stability and the semblance of life necessitate three fundamental properties ($TFP$). Firstly, these systems must possess the remarkable ability to engage in self-reproduction, enabling the potential transmission of their intrinsic characteristics or information to subsequent generations. This property engenders persistence and the propagation of the system within its environment.
Secondly, the structure of these chemical systems must exhibit the capacity for variation. This variability allows for the exploration of diverse possibilities, encompassing mechanisms such as mutation, recombination, or other forms of structural rearrangements. Through structural variation, these systems become adept at manifesting new traits or properties, facilitating their adaptation to the ever-changing circumstances of their existence.
Lastly, the maintenance of a far-from-equilibrium state assumes paramount importance. A continuous supply of energy is required to sustain the system's dynamic behavior. This energy input acts as the driving force, propelling the system away from thermodynamic equilibrium and unveiling a tapestry of complex, non-equilibrium behaviors.

 According to the statement above, a chemical compound, referred to as $\gamma$, has the potential to foster the development of a cognitive system within itself given favorable environmental conditions. When exposed to suitable energy sources, it is hypothesized that $\gamma$ can undergo a series of chemical reactions, leading to the formation of complex structures capable of processing information. Over time, these structures may evolve and become increasingly sophisticated, ultimately giving rise to a rudimentary cognitive system as $f(\gamma)$.
Thus, the cognitive mechanism, $f(\gamma)$,  of this particular type of chemical component, $\gamma$, is built upon $TFP$  as self-reproduction, structural variation, and maintenance in a far-from-equilibrium state. These properties give rise to characteristics reminiscent of cognitive processes, including information transmission, adaptation, and responsiveness to environmental stimuli.

Now consider $\gamma$ as a chemical compound with the capacity to influence cognitive mechanisms, represented by the function $f(\gamma)$. As $\gamma$ undergoes changes and evolves, we can interpret it as a representation of the brain, while the corresponding function $f(\gamma)$ can be seen as a representation of the mind, without loss of generality.
On the other hand, by drawing an analogy between the chemical compound $\gamma$ and its cognitive mechanism ability function $f(\gamma)$, we can establish a parallel to the relationship between the brain and the mind. In this analogy, the brain corresponds to the chemical compound $\gamma$, while the mind corresponds to the cognitive mechanism ability function $f(\gamma)$.

Pascal and Pross (2022)\cite{23pa}
 proposes that the ideas presented support a comprehensive approach to understanding the emergence of life. It suggests that life originates through contingent events within a specific context, where kinetic forces drive the development of more efficient self-reproducing systems. However, this process is constrained by the laws of thermodynamics and the properties of covalent bonds, particularly those involving carbon.
The paper emphasizes the essential role of organic chemistry, represented by $\gamma$, in the origin of life process. It highlights the influence of kinetic barriers associated with covalent bonds. Encouragingly, recent experiments have demonstrated that simple organic compounds can exhibit complex kinetic behavior, represented by $f(\gamma)$.
According to the paper's viewpoint, starting with the hypothesis of an auto-organizational process based on the kinetic properties of self-reproducing entities allows for a semi-quantitative assessment of the environmental conditions necessary for self-organization through organic chemistry. Notably, this assessment is consistent with visible light as an energy source and moderate temperatures.

In their (2023)\cite{Azar1}  paper, Azar introduced a topology for the system of the human mind, outlining various sets that contribute to its functioning. These sets are as follows:

$X_1$: Sensation set - Represents the reactivity of molecular reactions associated with sensory input.

$X_2$: Primary consciousness set - Involves the initial translation and processing of input data within the mind.

$X_3$: Awareness set - Encompasses knowing, perceiving, and being cognizant of events or stimuli.

$X_4$: Analyzing set - Involves the review and analysis of data within the mind.

$X_5$: Memory set - Represents the storage and retrieval of information and experiences.

$X_6$: Character mentality and mood set - Includes emotions, moods, and character traits that influence the mind's state.

$X_7$: System components coordinator set - Serves as the central part of the system, facilitating communication and coordination among various components.

$X_8$: Quality of will to accomplish an output set - Refers to the determination and motivation to achieve desired outcomes.

$X_9$: Curiosity along with perception set - Involves a deep sense of curiosity and exploration of phenomena or concepts without immediate analysis.

$X_{10}$: Other molecular, physiological, chemical, and physical conditions set - Represents additional factors related to the molecular, physiological, chemical, and physical aspects that influence the mind.

According that manuscript, these sets provide a framework for understanding the different aspects and components involved in the functioning of the human mind.

In that paper, the aforementioned components have emerged as a result of the evolution of the $\gamma$ system.
The proposed a topology for the mind endeavors to elucidate a discernible arrangement of system components ($X_i$ where $1\leq i\leq 10$) that encapsulates the functioning of a system resembling that of a human. Nevertheless, it is important to acknowledge that simpler entities such as artificial intelligence, plants, or insects may not possess the entirety of the aforementioned components.
These aforementioned components can be aptly described as $\gamma$ arms, which have gradually manifested over an extensive duration. These formations are the result of internal chemical and physical metamorphoses transpiring within the $\gamma$, in addition to the impact of external environmental factors.

Consider the existence of a set of sensitivities within the realm of the $\gamma$ system, represented by $X_1$. As time unfolds, the activities of $X_1$ give rise to the emergence of secondary sensitivities within the intricate fabric of the $\gamma$ system. These newfound sensitivities, collectively referred to as $X_2$, exhibit a notable responsiveness to the grouping of elements in $X_1$. Each element in $X_2$ manifests a distinct sensitivity to an individual element or a collective composition of elements from $X_1$, thus displaying a spectrum of sensitivities.
The genesis of $X_2$ can primarily be attributed to the molecular sensitivities inherent in the essence of $\gamma$. These sensitivities, both chemical and physical in nature, react and respond to fluctuations in the states of $X_1$.  It is important to note that, beyond its receptiveness to the external environment, $X_1$ also demonstrates a reciprocal responsiveness to internal processes, including those of $X_2$,  thereby establishing an ongoing cycle. This cyclic process paves the way for continual enhancement of the sensitivities exhibited by both $X_1$ and $X_2$,  ultimately leading to their refinement and development.

 As the process continues, a new generation of sensitizers are generated which are sensitive to the performance and changes of $X_2$. An example of these sensitizers can be represented by a category denoted as $X_3$. It is important to note that all of these sensitivities are molecular, chemical, and physical in nature, occurring within $\gamma$.
The elements of $X_3$ exhibit functional reactivity that depends on the changes of $X_2$ elements. 
It is worth noting that in the process of $\gamma$ evolution, other components such as $X_3, X_4, X_5$, etc., may arise in parallel as subsequent sensors within $\gamma$. Consequently, the system components, including the sets $X_1, \cdots, X_{10}$, develop a unique sensitivity to their constituent elements, which is refined over time. As the process continues, a one-to-one correspondence is established between the system components. For instance, if we consider $\gamma$ to be the brain, different frequencies of light received by $X_1$ evoke varying and corresponding sensitivities in $X_2$, which we perceive (or experience) as different colors.
 At this juncture, the chemical composition $\gamma$ may appear to be a simple system, and we refer to it as such. The various sensors that give rise to the system's components ($X_i$, where $1\leq i\leq 10$) are collectively referred to as "advanced sensors of all generations" ($ASAG$), or simply "sensors" according to Azar's paper (2023)\cite{Azar1}. It is important to note that these sensitivities develop incrementally due to a combination of environmental factors external to the system and chemical and physical reactions within the system.

The root of the experience of consciousness is essentially the result of the function of $ASAG$  which is described in detail in  Azar (2023)\cite{Azar1}.
It is important to note that the interface between a system and its environment is facilitated by sensors. These sensors enable the system to establish a correspondence between its experiences and the changes and events that occur in nature.
In this process, the experience of a system is dependent on the surrounding changes and the behavior of its sensors. It is crucial to evaluate the reality of a system's experience within its own context, as the perception of reality can vary between different systems due to their diverse sensors and components. For instance, most animals and insects are incapable of perceiving the roundness of the Earth or its movement around the Sun because their sensors lack the necessary capabilities. Similarly, humans face similar limitations in comprehending certain phenomena.
Our experience of light frequencies, for example, manifests as the observation of different colors. This phenomenon is directly related to the reactions of sensors in components such as $X_1$ and $X_2$. If our sensors operated differently, our experience of perceiving light would be altered accordingly. This principle holds true for other types of experiences as well.
In summary, the interface between a system and its environment is established through sensory mechanisms. The system's experiences are influenced by the surrounding changes and the behavior of its sensors. The perception of reality must be evaluated within the context of each system, as different systems possess varying sensors and components. This leads to different interpretations and understandings of the surrounding world. Our experience of phenomena, including light frequencies, is shaped by the reactions of specific sensors.

In the evolution of the cognitive mechanism, there is no clear demarcation point that separates the transition from non-living to living organisms (Azar (2023)\cite{Azar})  or from a state without consciousness to a state with consciousness (Azar (2023)\cite{Azar1, Azar2}). Instead, it is generally considered to be a gradual and continuous process.
The evolution of consciousness is believed to have occurred gradually over an extended period. It is not attributed to a single defining moment or specific genetic event but rather a cumulative result of biological and cognitive changes over time. Simple organisms possessed basic sensory and reactive capacities, which gradually developed into more sophisticated cognitive abilities in complex organisms.

The fundamental elements of sensory experiences are qualia, which refer to qualitative sensations or sense data. Qualia are considered the smallest units of phenomenal experience. The mind utilizes these units to form a detailed representation of the external world. The accuracy of these representations determines the extent to which they correspond to the actual external world, thereby constituting veridical perceptions. These states acquire their syntactic properties by virtue of their manifestation in the brain.
In other words,  qualia, or qualitative sensations, refer to the subjective aspects of our sensory experiences. They encompass the raw, immediate qualities of perception, such as the redness of an apple, the sweetness of sugar, or the warmth of sunlight on our skin. Qualia are highly personal and subjective, as they are experienced individually and cannot be directly shared or communicated with others. They are considered the basic building blocks of our conscious experiences.
In every scientific discipline, a consistent set of subject axioms and methods of reasoning is necessary to address the questions and challenges within that field. If we consider qualia as the foundational axioms for the set of conscious experiences, then the definition of these axioms requires additional fundamental axioms that maintain consistency within the system.
When analyzing qualia, we rely on the conscious experiences of the system, which are derived from qualia itself. As a result, certain questions related to the thesis, such as defining the experience of consciousness within philosophical frameworks, become difficult to answer in isolation. It becomes necessary to integrate insights from other scientific disciplines such as biology, mathematics, physics, or chemistry.
By combining knowledge from these diverse fields, we can gain a more comprehensive understanding of the nature of consciousness and develop a systematic analysis that takes into account various perspectives and empirical evidence. This interdisciplinary approach is crucial to tackle the complex questions posed by qualia and to ensure a more holistic understanding of conscious experiences.

\section{Mathematical Approaches in the Study of Consciousness Experience}

Integrated Information Theory (IIT) has gained a lot of attention for potentially explaining, fundamentally, what is the physical substrate of consciousness.
According to the IIT, consciousness is believed to emerge from the integration of information across a highly interconnected network of elements. The theory suggests that the extent of consciousness is directly related to the quantity of information integrated within a given system. 
  Tononi  (2008)\cite{27t} has been studied
Integrated information  and he defined the function $\phi$
as the amount of information generated by a complex of
elements. 
The foundational concepts behind IIT were extremely innovative, and it has been very exciting to see certain predictions being upheld in experiments. 
It aims to describe both the quality and quantity of the conscious
experience of a physical system, such as the brain, in a particular state.
An IIT aims to specify for each system in a particular state its conscious experience. As such, it will require a mathematical model of such experiences.
 Kleiner and Tull (2020)\cite{7kt} aimed to provide a comprehensive description of both the quality and quantity of conscious experiences within a physical system, such as the brain, in a specific state. Their contribution focused on presenting the mathematical framework of the theory, distinguishing the fundamental elements from auxiliary formal tools. Additionally, they put forth a definition of a generalized Integrated Information Theory (IIT).
Numerous books and articles, including Chalmers (2003\cite{7c}, 2009\cite{8c})) and others, have been dedicated to the exploration and examination of consciousness, as well as the quest for an exact definition of this elusive phenomenon. These works delve into the intricacies of consciousness, exploring its nature, properties, and theoretical frameworks in an attempt to shed light on its fundamental aspects.
Chalmers presents the hard problem of consciousness, which refers to the challenge of explaining how and why subjective experiences arise from physical processes in the brain. He argues that purely physical explanations, such as those based on neuroscience, fail to account for the subjective aspect of conscious experience.
Chalmers distinguishes between the "easy problems" of consciousness, which involve explaining cognitive functions like perception and attention, and the hard problem, which focuses on the subjective experience itself. He suggests that a satisfactory theory of consciousness must bridge the explanatory gap between physical processes and subjective experience.

The question of defining consciousness and understanding its nature has been a subject of deep contemplation and debate among philosophers, neuroscientists, and cognitive scientists. While there is no universally agreed-upon definition of consciousness, it generally refers to the subjective experience of being aware of oneself and the surrounding world.
 Philosophers often ponder the question of whether the concept of consciousness can be precisely defined. If we entertain the idea that a comprehensive definition of consciousness is attainable, it would inherently be an experience of consciousness in itself. Furthermore, if such a definition were to be achieved, it would need to be capable of encompassing and expressing the entirety of our experiences through its inherent characteristics.

One of the challenges in defining consciousness precisely is that it is an experiential phenomenon. Any attempt to define consciousness would inherently involve experiencing it, which creates a circularity in the definition. In other words, in order to define consciousness comprehensively, one would need to have a conscious experience of that definition.
Moreover, if we assume that a comprehensive definition of consciousness is achievable, it would need to encompass and express the entirety of our subjective experiences. Consciousness is not limited to a single aspect or isolated phenomenon but encompasses a wide range of mental states, perceptions, thoughts, emotions, and sensory experiences. Any definition that aims to capture the full scope of consciousness would need to account for these diverse aspects.

When considering this all-encompassing definition, it implies that various phenomena can be categorized as experiences. Let's denote the set of such phenomena as $X=\{A\notin A: A~\text{is a set}\}$. It becomes evident that $\{1\}$ and $\{1,2\}$ belong to $X$, suggesting that $X$ is not empty. Now, viewing $X$ as an experience, we raise the question of whether $X\in X$ or $X\notin X$. Logically, this question should have an answer, yet none can be found. Through mathematical analysis, I demonstrate that it is not possible to define all of our experiences with specific characteristics.
Consciousness phenomena are not finite or quantifiable in terms of numbers. Instead, the transition from one phenomenon to another is a continuous process, demonstrating a cardinality comparable to that of real numbers.

In the following, we demonstrate that the occurrence of all physical signals in the brain is unbounded. It is known that for a conscious experience in the mind, there is at least one physical signal that occurs in the brain. Let $\mathcal{P}$ represent the set of all physical signals occurring in the brain, and $\mathcal{E}$ represent the set of conscious experiences in the mind. Thus, there exists a function $\Gamma$ from $\mathcal{P}$ onto $\mathcal{E}$ such that for any conscious experience $Y \in \mathcal{E}$, there exists $X \in \mathcal{P}$ satisfying $\Gamma(X) = Y$.
It is evident that the set of conscious experiences in the mind is unbounded. For instance, for each natural number $n$, there exists a conscious experience denoted as $f(n)$, and there exists a relationship $n \longleftrightarrow f(n)$ between them. Since the set of natural numbers is infinite, the set of conscious experiences derived from numbers is also infinite. Thus, the occurrence of all physical signals in the brain is unbounded.
This outcome indicates that the information stored in our memory is not constrained. Consequently, the processing and analysis of information in our minds will not be limited.

According to the given information, we can establish a relationship between the set of conscious experiences denoted as $\mathcal{E}$ and the corresponding physical processes in the brain represented by $\mathcal{P}$. By categorizing the elements of $\mathcal{P}$ that contribute to specific experiences, we can define a function $\Gamma^{-1}$ that allows us to trace back from an experience in $\mathcal{E}$ to the associated elements in $\mathcal{P},$ enabling a deeper understanding of the underlying processes involved.
The set $\Gamma^{-1}(\mathcal{E}) = \{\Gamma^{-1}(Y): Y \in \mathcal{E}\}$ represents the complete physical domain of the brain in relation to the set of conscious experiences. However, it is important to acknowledge that not all physical signals in the brain lead to conscious experiences. Consequently, $\Gamma^{-1}(\mathcal{E})$ is a subset of $\mathcal{P}$, and it does not encompass the entire set $\mathcal{P}$. To avoid ambiguity, we restrict the function $\Gamma$ to its domain, and we assume that the domain of $\Gamma$, denoted as Domain$\Gamma$, is equal to $\mathcal{P}$. Therefore, we can assume $\Gamma^{-1}(\mathcal{E}) = \mathcal{P}$.
This implies that there exists a one-to-one correspondence between the elements in $\mathcal{P}$ and $\mathcal{E}$.  For instance, if $\alpha$ symbolizes a signal interpreted as a light frequency, it undergoes successive processing steps $X_1$ and $X_2$, leading to the perception of a distinct color. In this context, $\Gamma^{-1}(X_2(X_1(\alpha)))$ represents the physical processes occurring in the brain that are linked to the perception of that specific color.

   The collection of all experiences of a complete system cannot be a set, as it would lead to a contradiction in set theory. Assuming that $\mathcal{E}$  is the set of all experiences of a complete system, we would have $\mathcal{E} \in \mathcal{E}$, which contradicts the axioms of set theory.
To address this issue, we can consider an alternative approach for the collection of system experiences. If we view all humans as a system, we know that each individual human has a finite number of experiences within a given time interval $[t_1, t_2]$. If we define $\mathcal{E}(t)$ as the collection of all experiences  up to time $t$, then $\mathcal{E}(t)$ is finite.
It becomes evident that $\mathcal{E}(t_1) \subseteq \mathcal{E}(t_2)$ whenever $t_1 \leq t_2$. We can then define $\mathcal{E}$ as the union of all $\mathcal{E}(t)$, denoted as $\mathcal{E} = \bigcup_t \mathcal{E}(t)$. This implies that $\mathcal{E}$ is a countable set representing the possible experiences of the system.
By adopting this approach, we avoid the contradiction arising from considering the collection of all experiences as a single set and instead recognize that experiences can be organized and categorized within a temporal framework, leading to a countable set of potential experiences for the system.

 The approach you're describing involves defining a hierarchy of sets $A_0, A_1, A_2, \ldots$ to represent different levels of experiences. The set $A_0$ consists of experiences that do not result from other experiences, such as seeing colors, feeling pain, or hearing sounds  which are called qualia or subjective experiences.   The set $A_1$ represents experiences that result from the experiences in $A_0$, and $A_n$ result from the experiences in $A_0, A_1,\ldots, A_{n-1} $.
However, when we define $\mathcal{E} = \cup_{i=1}^{+\infty}A_i$, we encounter a difficulty in treating $\mathcal{E}$ as a set. This is because the process of taking the union of infinitely many sets may not necessarily result in a set according to the axioms of set theory.
This issue is related to the concept of a "set of all sets," which leads to paradoxes in set theory, such as Russell's paradox. The paradox arises when considering whether a set can contain itself as an element. In our case, $\mathcal{E}$ would contain all sets $A_n$ as its elements, including itself, which leads to a contradiction.
To avoid this paradox, one possible approach is to consider the hierarchy of sets $A_0, A_1, A_2, \ldots$ as a potential progression of experiences rather than a set-theoretic construction. In this view, $\mathcal{E}$ would represent the collection of all possible experiences, but it would not be treated as a set in the traditional sense. Instead, it would be an abstract concept denoting the entirety of possible experiences within the system.
By reframing $\mathcal{E}$ as a concept rather than a set, we can still discuss the progression of experiences and their relationships without running into the paradoxes associated with the set of all sets.

Assume that a function  $\Gamma$  processes of our brain's behavior to a set of experiences $\mathcal{E}$. On the other hand the function $\Gamma^{-1}$ will be from $\mathcal{E}$ into all of our brain's behavior. By restricting $\Gamma^{-1}$ to $A_0$, we obtain $\Gamma^{-1}(A_0)$, which represents the set of brain processes caused by external stimuli. Therefore, all activities and behaviors of the brain that result in an experience can be derived from the extension of the function $\Gamma^{-1}$ to the complete set of experiences, denoted as $\mathcal{E}$.
Additionally, based on the definition of component $X_2$, it is evident that $X_2 = A_0$. Consequently, if $\mathcal{E}$ is not a set, then $X_2$ is also not a set. The reason for this issue lies in the existence of different experiences originating from the same phenomenon. For instance, perceiving the color red with varying concentrations or imagining it in a dream can lead to distinct experiences.
In other words, the set of all possible experiences for a human system within a given time period can have an arbitrary size. To illustrate, let's consider representing the colors white and black as $w$ and $b$, respectively. We can define a function $f$ that maps the interval $[w, b]$ to the interval $[0, 1]$, where the elements of $[w, b]$ represent combinations of colors between white and black. Specifically, $f(x)$ denotes the degree of color indicating the proximity to white or black when $x \in [w, b]$. Function $f$ is both one-to-one and surjective. Furthermore, we define $d(x, y) = \vert f(x) - f(y) \vert$ for any $x, y \in [w, b]$, where $d$ represents the distance between $x$ and $y$ in terms of color perception.
It can be concluded that $f$ is an isomorphism. However, when considering the set of all possible experiences of seeing or imagining colors between white and black, from a mathematical standpoint, this set has a cardinality of $c$. It is important to note that while mathematically there may be an infinite number of possible experiences, not all of them are attainable or realized in practice. This realization contradicts the notion of $\mathcal{E}$, which is defined to be countable.

Let us now contemplate a scenario wherein our experiences undergo a continuous transition from a state of near-absolute brightness to one of absolute darkness, as exemplified by the observation of a sunset. In this particular context, it becomes apparent that claiming to have experienced the entirety of this period as a unified whole would be misleading. The passage of time, specifically from the stage of complete illumination to that of complete obscurity, assumes a pivotal role in shaping our experiences. The duration of this transitional phase can vary significantly, spanning from mere minutes to prolonged hours, or even extending across multiple days, thereby giving rise to diverse individual experiences. Put differently, the nature of our experiences, encompassing the perception of light gradually transforming into the perception of utter darkness, hinges upon the precise timing of this process.
In essence, the unique character of our subjective encounters is intimately intertwined with the temporal intricacies governing the transition from luminosity to shadow.

Hence, as we traverse the realm of our experiences, it becomes increasingly apparent that drawing a distinct boundary between darkness and light presents itself as a formidable endeavor. In simple terms, when we witness the gradual shift from luminosity to obscurity or vice versa, our perceptions do not lend themselves to establishing a definitive moment of demarcation. We find ourselves unable to pinpoint an exact point in time where we can confidently assert that prior to a specific moment "$t$" there existed light, and subsequent to that moment, darkness ensued. The fluidity of this transition defies our attempts to impose rigid divisions within the continuum of our perceptual journey.

Indeed, as we navigate through the realm of our experiences, it becomes increasingly apparent that delineating a clear boundary between darkness and light proves to be a formidable task. In simple terms, when we observe the gradual shift from luminosity to obscurity or vice versa, our perceptions do not readily lend themselves to establishing a definitive moment of demarcation. We find ourselves unable to pinpoint an exact point in time where we can confidently assert that prior to a specific moment "$t$" there existed light, and subsequent to that moment, darkness ensued. The fluidity of this transition defies our attempts to impose rigid divisions within the continuum of our perceptual journey.

 Within the realm of philosophical and scientific inquiry, a multitude of scholars and thinkers embark on the quest to unravel the intricacies of human experiences through the formulation of rules and methodologies. However, a fundamental limitation emerges when endeavoring to comprehensively analyze a system from within itself and establish overarching governing principles.
For the sake of argument, let us assume the existence of a comprehensive framework designed to analyze and understand all experiences within a specific system. We denote this framework as $\mathcal{E}$. Moreover, let us consider that $\mathcal{E}$ represents a subset of the set encompassing all analyzable experiences, denoted as $\mathcal{E}$.

 Let's consider $\mathcal{E}$ to represent an individual experience. In this context, we can define the set $\mathcal{H}$ as the collection of times denoted by $t$ when $\mathcal{E}$ does not occur, which we represent as $\mathcal{E}(t)$. Mathematically, we can express this as follows:
\begin{align*}
\mathcal{H}=\{t:~\mathcal{E}\notin \mathcal{E} (t)\}.
\end{align*}
It is clear that $\mathcal{H}$ is a non-empty set. Since $\mathcal{H}$ is bounded above, we can establish its supremum as $t_0$. Consequently, for any $t>t_0$, $\mathcal{E}$ belongs to $\mathcal{E}(t)$. Furthermore, the creation of $\mathcal{E}$ arises from the collection of all our experiences at time $t_0$, denoted as $\mathcal{E}(t_0)$. Assuming that $\mathcal{E}$ can analyze $\mathcal{E}$, $\mathcal{E}(t)$ can produce $\mathcal{E}$ for all $t>t_0$. However, $\mathcal{E}(s)$ cannot produce $\mathcal{E}$ for all $s<t_0$.

Now, let's consider $s<t_0<t$. When examining the time interval $[s,t]$, we notice that our experiences will be severely limited when $t$ and $s$ are extremely close to each other. This leads to a contradiction in our assumption.

\section{Extracting Logical and Conceptual Phenomena from Experiential Realms}
Living organisms are formed through an ongoing process of chemical and physical interactions involving specific chemical compounds. These compounds have evolved within cognitive systems, which possess distinct characteristics within the system of living organisms. This information is supported by the article "Azar (2023)\cite{Azar}" and its associated sources.

Cognitive systems undergo a transformative process, transitioning between chemical and biological states, encompassing both non-living and living entities. These systems consist of a fusion of chemical sensors,  $ASAG$ (Azar (2023)\cite{Azar1, Azar2}), which emerge through the interplay of assessing the chemical composition of the external environment and internal chemical and physical processes. As a chemical compound evolves into a living organism, the cognitive mechanism responsible for this transformation is further amplified and refined.
According to Pamela et al. (2021)\cite{Pamela}, basal cognition refers to the essential processes and mechanisms that enabled organisms to monitor specific environmental conditions and respond adaptively to ensure their survival (e.g., finding food, avoiding danger) and reproductive success long before the emergence of complex nervous systems, let alone central nervous systems. Basal cognition implies an inherent level of implicit familiarity or comprehension of the connections between environmental states.
In the pursuit of enhanced thermodynamic stability, which forms the foundation for all physical and chemical processes, one of the primary avenues is through the well-established principles of thermodynamics.
 Azar (2023)\cite{Azar} posits that distinguishing between living and non-living organisms  based on their characteristics is not feasible. The transition from a non-living chemical compound to a living entity is a continuous process without a clear demarcation point.
In their manuscript (Pascal and Pross, 2023)\cite{23pa}, the authors propose that perception arises from the dynamic interplay between the replicative system known as the Dynamic Kinetic Stability (DKS) system and its supportive environment. The DKS system is considered a kinetic phenomenon associated with the self-organization of life, driven by the power of replication.
Pascal and Pross argue that the DKS system, through kinetic selection, "learns" which variations are advantageous and which are not. This capacity for replication allows the system to both remember and learn, indicating a mental dimension. Consequently, the concept of DKS provides a framework for understanding the physico-chemical basis of the mental aspects of life.

When a system arises due to the interplay of physical and chemical factors in its environment, the logic and regularity observed within the system are closely intertwined with the underlying physical and chemical properties of that environment. In simpler terms, the behavior and functioning of the system's components are directly influenced by the mobility and characteristics of the surrounding environment, as well as the internal chemical and physical reactions of the system that interact with it.
As the system interacts with its environment, it becomes attuned to the patterns and cues present within its surroundings, perceiving a certain degree of regularity.
This perception of regularity enables the system to exhibit intelligent adaptation and responsiveness based on the available environmental cues. Through the interplay between the system and its environment, the system can discern meaningful information, recognize recurring patterns, and adjust its behaviors or internal states accordingly. This adaptation allows the system to optimize its functioning and enhance its chances of survival or achieving its objectives within the given environmental context.
It is important to note that the notion of intelligent adaptation and perception of regularity does not necessarily imply conscious awareness or intentionality. Instead, it reflects the inherent interconnectedness between the system and its environment, where the system's behavior emerges from the dynamic interactions between its components and the external factors at play.
This perspective acknowledges the profound influence that the physical and chemical properties of the environment exert on the behavior and functioning of systems, highlighting the intricate relationship between a system's internal processes and its external context.

The regularity manifested within a system can be comprehended as a correspondence or equivalency between the internal processes intrinsic to the system and the performance exhibited by its encompassing environment. Given that the genesis and operation of a system are subject to the influence of the chemical and physical properties characterizing its environment, a correlation emerges between the activities and mobility of the system itself and the activities transpiring within the ambient milieu.
Put simply, when an individual discerns what may appear as intelligent deliberation or purposeful reasoning in their immediate surroundings, it ensues as a consequence of the system's internal adaptation to the physical and chemical functions inherent to the environment. The correspondence between the system and its chemical and physical environment consequently engenders the perception of orderly regularity within our environment, which may inadvertently foster the supposition of intentional planning.
To expound upon this matter with greater precision, the intelligent mechanisms observed in natural phenomena are indeed manifestations of the system's correspondence with its surrounding environment. They do not serve as indications of deliberate planning or external intelligence. Rather, they arise from the intricate interplay between the system and its environment, whereby the system's behavior and adaptive qualities emerge from its inherent properties and the influences exerted by the encompassing physical and chemical factors.

The genesis of a system is intricately tied to the intricate web of chemical and physical interactions unfolding within its surrounding environment. This mode of emergence engenders a profound correspondence between the system and its environment, setting the stage for the system's perception and conceptualization of regularity, which may present itself as exhibiting intelligent reasoning. However, it is essential to recognize that the true elucidation lies in the inherent correspondence or equivalence characterizing the system and its environment.
To illustrate this idea, let us consider a hypothetical scenario wherein we construct an intelligent machine endowed with the capacity to assess human behavior. As this machine observes and analyzes our actions, it discerns a discernible regularity and logical progression, seemingly evoking intelligent deliberation. Yet, this apparent intelligence arises from the approximate correspondence or equivalence established between our behavior and the internal workings of the machine itself. This correspondence is established during the construction and design of the machine, grounding its perception of regularity and intelligent thought.
Consequently, as the system interacts with its environment over time, undergoing a series of physical and chemical reactions, an equivalence relationship emerges between the functioning of the system and the intricate dynamics of its surrounding milieu. It is within this relationship of correspondence that the system's perception of regularity is shaped and the semblance of intelligent thought may manifest.
In essence, the interplay of physical and chemical processes transpiring in the system's environment engenders a profound equivalence with the system's functioning. This resulting correspondence not only shapes the system's perception of regularity but also gives rise to the appearance of intelligent thought, offering insights into the intricate relationship between a system and its environment.

 The order, regularity, and logical patterns that a system perceives in its surrounding environment are a result of the equivalence relationship between the system and its environment. This implies that there is no inherent regularity or logic in nature itself, and the perceived regularity by a system is a product of this equivalence.
The foundation of this equivalence relationship lies in the way systems are created. As explained earlier, the sensitivities of a system are based on the physical and chemical behaviors of its environment. Changes in the environment can trigger responses in the system, making it sensitive to those changes. Some systems have sensors that can record these changes in their memory, allowing them to compare and recognize similar changes when detected by their sensors.
In essence, the system's sensitivity to its environment and its ability to record and compare changes contribute to the perception of order, regularity, and logical conclusions within the system. This perception arises from the equivalence established between the system and its surrounding environment.

In our daily lives, the logic we use serves as a cognitive tool that prioritizes coherence and minimizes contradictions. Initially, we learn this logic by observing nature, and then we refine and develop it through conscious experiential processes. It is important to note that our focus is not on delving into the complexities of mathematical or philosophical logic, but rather on understanding how humans construct logical frameworks for themselves.

According to the Section 2, we know that the  dynamic kinetic stability and the semblance of life are reliant on three fundamental properties ($TFP$).
The manifestation of characteristic $TFP$ in living organisms, spanning from single-celled organisms to humans, is attributed to the function of $ASAG$. The evolutionary progression of sensors has consequently influenced the development of $TFP$, thereby enhancing the capabilities of sensors within the systems we encounter. As a logical outcome, the processes occurring within the system are a consequence of the chemical and physical activities involved in the creation and evolution of $TFP$.

The presence of system consciousness enriches and reinforces this logical process, aligning its performance with the enhancement and optimization of $TFP$. The creation and evolution of conscious experience give rise to another logical process that emerges from the system's accumulated experiences. Consequently, the logical behavior observed in primitive living organisms lacking animal consciousness experiences can be characterized as pseudo-logical, as it is limited to the maintenance and promotion of $TFP$ without the additional depth and complexity brought about by conscious awareness.

The advancement of the system's logical process occurs concurrently with the development of the system's conscious experience. As sensors undergo upgrading or the system evolves along the path of $TFP$ evolution, other sensors such as intelligence, talent, emotion, and a sense of responsibility are created.
The existing logic within the system is a consequence of the evolutionary process of the sensors ($ASAG$) present in the system. This logic functions to uphold and progress $TFP$. We denote this quasi-logical behavior as $L_{TFP}$.

The establishment of this logic is a common outcome across various living organisms, arising from the interplay between their internal system and the behavior of their surrounding environment.
In this context, "correspondence" refers to the chemical and physical reactions occurring within the system, which are directly influenced by the behavior of the surrounding environment. In essence, the system's activity is a function of the behavior exhibited by its environment.
Put differently, the chemical and physical changes taking place within the systems have established a correspondence with their surroundings, serving to maintain and stabilize the state of $TFP$, according to the paper of  Pross (2021)\cite{23pp}. Consequently, this has contributed to the emergence of a molecular and chemical-physical logic (or quasi-logic) within the systems, driven by various natural forces.
Indeed, the interaction of forces and the establishment of a unique state on Earth and in the solar system have played a role in initiating chemical and physical processes within the system, ultimately leading to the creation of $L_{TFP}$ within the system.

The development of human conscious experience is facilitated by the emergence and activation of new sensors within the system, see Azar (2023)\cite{Azar1}. As the process of complete system consciousness unfolds, it gives rise to pseudo-logical behavior ($L_{TFP}$), which can lead to illusions, heightened skills, and other abilities in individuals.
For instance, phenomena such as love, falling in love, or experiencing pleasure from food can be attributed to the development of the $L_{TFP}$ mechanism. However, the development of new sensors can also give rise to illusions in humans and certain animals, sometimes leading to challenges in their relationship with $L_{TFP}$.
Examples of these challenges include instances of humans engaging in harmful actions like killing children or self-starvation due to delusional beliefs, as well as cases of animals exhibiting self-destructive behaviors. These actions can be attributed to the activity of newly developed sensors.
Throughout history, the occurrence of illusions in humans has increased, and this can be directly linked to the expansion of human consciousness 
potential specifically, the birth of new sensors as part of the $ASAG$ system.
In this context, an illusion refers to an emotional response to an event or internal thoughts and feelings that are largely unrelated to the event itself or our true desires. For instance, when a person encounters an internal or external event, they may attribute an emotional cause to it that is not the primary cause of the event. Furthermore, they may propose a solution to address this emotional cause that is completely unrelated to the actual event.

The creation of new sensors within the human system leads to the formation of additional mental images. These new images in the mind are a consequence of upgrading the system's sensors ($ASAG$), or more broadly, upgrading the system's components. This process has resulted in an improvement in the human consciousness system.
As mentioned earlier, the evolution of conscious experience in humans has given rise to certain illusions, many of which contradict $L_{TFP}$. These illusions arise due to the development of our sensors, or in other words, the development of our conscious experience, causing us to act beyond the bounds of $L_{TFP}$.
The primary cause of these illusions can be attributed to the tendency of sensors to create illusions when receiving messages. These messages can take the form of fears related to events or internal aspects, dreams experienced during sleep or wakefulness, or specific feelings generated within the system by the sensors.
At this stage, the human consciousness system continually produces illusions, and individuals live their lives based on these illusions. In some cases, individuals seek to enhance the status of $TFP$, and this enhancement is often based on skills and illusions. Unfortunately, as previously mentioned, negative outcomes can arise from this pursuit.
However, it is important to note that during this stage, not only are illusions created, but human skills and abilities also grow.
In this context, illusions refer to instances where a person attributes a reason or explanation to an experience that has little or no connection to the actual cause of that experience.
The main cause of these illusions in humans at this stage is often due to the weaknesses of the system's components, particularly the weakness of the $X_4$ component.
In some cases, illusions are passed down to new generations by their predecessors, and intriguingly, these individuals may possess illusions that are not effectively scrutinized by the $X_4$ and other components.
It is worth noting that while problems or deficiencies in system components or illusion training can contribute to the creation of illusions, our discussion here focuses on the natural and normal conditions that give rise to illusions.

The development of human consciousness is intricately linked to the expansion of social life, which, in turn, gives rise to the emergence of skills and the evolution of components within the human system. These developments contribute to the advancement of consciousness.
Over time, a specific stage of sensor development occurs that is unique to humans. Consequently, in conjunction with the development of skills and empirical knowledge, the analysis component $X_4$ undergoes development and begins to analyze illusions.
This process of analysis and skill improvement has led to the establishment of a logical framework that seeks to identify the exact causes behind various phenomena. At this stage, the acquisition of logical reasoning by humans hinges on understanding the behaviors of nature and the relationships between different forces and laws of nature, both in relation to each other and to humans.
The enhancement of our sensors or system components enables us to identify the precise cause of most events. By repeatedly experiencing and observing these events, we come to grasp the true cause-and-effect relationships that govern them.
As an example, at this stage, individuals learn that earthquakes are caused by the movement of underground rocks along fault lines, rather than being attributed to the anger of unseen forces or entities.
This process leads to valid arguments such as
\begin{enumerate}
\item $(p_0 \wedge
\forall n (1\leq n\leq m) (p_{n-1}\Rightarrow p_{n}))\Rightarrow p_m$,
\item  $(p \wedge \sim p  )\Rightarrow q$,
\item  $p\Rightarrow (q \vee \sim q)$,
\item  $\forall x P(x)\Rightarrow \exists x P(x)$,
\end{enumerate}
 more detail see MacFarlane (2021)\cite{21m}.
In the first inference of the propositions mentioned above, $(1)$, our assumption is based on $p_0$. It is important to establish the validity of $p_0$ and ensure that it is not derived from other propositions or dependent on other statements. This raises the question of how we can be confident that $p_0$ is indeed valid.

In the realm of logical reasoning, there are certain propositions that are considered axioms. These propositions are typically self-evident or universally accepted and serve as the foundation for further reasoning. They are not derived from other propositions but are treated as valid starting points.
For instance, propositions that are equivalent to a valid statement such as $p \vee \sim p$ (the law of excluded middle) are considered axioms. These propositions are inherently valid based on their logical structure.
On the other hand, we cannot assert with certainty that an event will or will not occur in a specific moment. Therefore, any proposition that leads to a contradiction, such as $p \wedge \sim p$, is considered invalid.
In summary, in order to establish the validity of a proposition like $p_0$, we need to examine its logical structure, determine if it is an axiom or derived from axioms, and ensure that it does not lead to a contradiction. By adhering to these principles of logical reasoning, we can evaluate the validity of propositions and construct a reliable framework for knowledge and inference.

Indeed, many people rely on the simple logical structure outlined earlier to assess the correctness and falsity of propositions and to avoid falling into illusions. This logical framework, which includes principles such as the law of excluded middle and the law of non-contradiction, is widely accepted and serves as a basis for reasoning.
However, it is important to note that some individuals may possess knowledge of logic but still entertain delusional thoughts. In such cases, factors such as the person's mental state, incorrect education, or individual personality traits can contribute to their distorted thinking.
The discussion here primarily focuses on the evolution of human consciousness and how it relates to understanding the truth or falsity of propositions. Discerning the truth or falsity of a proposition is indeed a conscious experience, and our understanding of it is influenced by various factors.
When dealing with the truth or falsity of a proposition, we encounter several other considerations. Firstly, we strive to ensure the certainty of logical validity, which involves evaluating the logical soundness of the proposition and its conformity to established principles. This step helps us establish a solid foundation for our reasoning.
In the second step of this process, we aim to assess the reliability or quality of our conscious experience. This involves examining the accuracy and consistency of our perception, cognition, and interpretation of the proposition and its associated information. By ensuring that our conscious experience meets certain standards, we can enhance our ability to discern the truth or falsity of propositions more effectively.
In general, a phenomenon can be classified as a logical phenomenon when it satisfies the following conditions:

$A_1$: The phenomenon is a result of natural forces operating in the world.

$A_2$: It is a logical consequence of the underlying natural processes described by $A_1$.

$A_3$: The representation of this phenomenon in our conscious experience is both accurate and comprehensive.

By meeting these criteria, a phenomenon can be considered as a logical phenomenon, indicating that it adheres to the principles of logic and is consistent with our understanding of the natural world.
Indeed, our ability to engage in logical reasoning and our conscious experience find their ultimate foundation in the intricate chemical and physical processes that occur in nature, including those within our own bodies and brains. These underlying processes play a pivotal role in shaping various aspects of our cognition, perception, and the functioning of neural networks.
Within our bodies, complex chemical reactions and interactions take place, enabling the transmission of signals between neurons and the formation of synaptic connections. These processes contribute to the establishment and modulation of neural networks that underlie our cognitive abilities, including logical thinking.
Furthermore, our perception of the world and our conscious experiences are shaped by the way our sensory organs detect and process various stimuli, converting them into electrical signals that the brain interprets. The neural processes involved in perception, attention, memory, and higher-order cognitive functions are influenced by the underlying chemical and physical processes occurring within our brains.
By understanding the intricate interplay between these chemical and physical processes and their impact on our neural networks, we can gain insights into the mechanisms that govern our logical thinking and conscious experiences. This understanding highlights the intimate connection between the physical world, our bodies, and the cognitive processes that give rise to our capacity for logical reasoning and conscious awareness.

As a result, our logic is often based on natural phenomena that are readily observable and widely accepted as axioms or foundational principles. These phenomena serve as the building blocks of our logical frameworks and provide a basis for making inferences. For example, principles such as cause and effect, the conservation of energy, and the laws of mathematics are derived from our observations of the natural world.
By grounding our logic in these natural phenomena, we establish a logical framework that aligns with our understanding of the world and enables us to make reliable inferences. These foundational principles, often considered axioms, offer a solid foundation for constructing logical arguments and reasoning about various aspects of reality.
However, it is important to note that while our logic is influenced by natural phenomena, it is also subject to refinement, expansion, and revision through ongoing scientific inquiry and philosophical discourse. As our understanding of the natural world advances, our logical systems may evolve to accommodate new insights and discoveries.
In summary, our logic is fundamentally connected to the natural world and is influenced by observable phenomena. By recognizing and utilizing these natural phenomena as axioms, we can construct logical frameworks that facilitate accurate reasoning and inference. Nevertheless, it remains crucial to remain open to further exploration and refinement of our logical systems as we continue to deepen our understanding of the world.

Indeed, the logic that humans develop for themselves is shaped by a careful examination of nature's behavior and their own experiences. The advancement of sensors and awareness, often referred to as $ASAG$ promotion, plays a crucial role in facilitating this process. Our logic is primarily influenced by two main factors:
\begin{enumerate}
\item  Influence of Natural Forces,
\item Role of Conscious Experience.
\end{enumerate}

By engaging with these two fundamental influences, humans develop a logic that is informed by the forces of nature and the depth of conscious experience. This intertwined relationship between external observations and internal reflections allows for the continuous refinement and evolution of human logic.  

In the current stage of our development, our conscious experience has the capacity to enhance itself through the improvement of our sensory capabilities. This upgraded conscious experience strives to construct a logical framework that applies to both individuals and society, similar to the principles of the  $L_{TFP}$ that govern human and social systems. We can refer to this new logic as $L_{\widehat{TFP}}$, which serves as an extension of $L_{TFP}$.

The emergence of $L_{\widehat{TFP}}$ signifies an evolution in our logical framework, incorporating enhanced sensory input and an expanded understanding of the world. It allows for a more comprehensive and nuanced approach to reasoning, inference, and decision-making. By leveraging the advancements in our conscious experience and aligning them with the principles of $L_{TFP}$, $L_{\widehat{TFP}}$ aims to provide a more robust and adaptable logic for individuals and society as a whole.

The improvement of our sensors indeed plays a crucial role in the development of this logic. Consequently, $L_{\widehat{TFP}}$ expands and further enhances the framework provided by the pseudo-logic $L_{TFP}$.

The advancement of human consciousness, coupled with the analysis component $X_4$, contributes to the examination and analysis of delusions. When we observe or experience a phenomenon, we engage in a more precise investigation of its underlying causes, guided by the logical process of $L_{\widehat{TFP}}$.

Throughout this process, we strive to establish consistent axioms that explain phenomena through logical arguments, with the aim of minimizing contradictions and paradoxes within this logical framework. The goal is to construct a coherent and robust system that can effectively account for the complexities of the world while maintaining internal consistency.

By employing the logical tools provided by $L_{\widehat{TFP}}$ and utilizing the insights gained through the advancement of our conscious experience and sensory capabilities, we can refine our understanding of the world and address delusions or misconceptions that may arise. This ongoing process of examination and analysis helps to strengthen the logical foundations and promote a more accurate and comprehensive understanding of the phenomena we encounter.

Now consider the following three modes for the pseudo logic of living beings:
\begin{enumerate}
\item[$L_1$:] The process of maintaining and evolving the concept of $TFP$ can be described as a quasi-logical operation denoted as $L_{TFP}$.

\item[$L_2$:] A quasi-logical process utilized for the preservation and development of $TFP$, incorporating a blend of illusions and skills.

\item[$L_3$:] The logical process employed to uphold and enhance $TFP$ (alongside skills and meticulous event analysis) can be represented as $L_{\widehat{TFP}}$.
\end{enumerate}

In the framework of cognitive systems, the quasi-logical process denoted as $L_1$ serves as a mechanism for preserving the state of $TFP$. However, in the $L_2$ mode, living beings employ a quasi-logical process to maintain and evolve $TFP$ while incorporating illusions and skills. In this mode, beings combine logical reasoning with illusions, which can be subjective interpretations or distortions of reality, along with skills acquired through experience.
Moving beyond $L_2$, the $L_3$ mode represents a logical process employed to maintain and improve $TFP$. This mode encompasses both skills and detailed analysis of events. Referred to as $L_{\widehat{TFP}}$, it extends the previous modes by incorporating more sophisticated logical reasoning. In $L_3$, systems engage in a comprehensive analysis of events and phenomena, utilizing their skills alongside logical reasoning to enhance and refine $TFP$.

The three modes, $L_1$, $L_2$, and $L_3$, represent distinct levels of logical engagement and sophistication in the context of maintaining and evolving $TFP$. These modes denote different stages of logical development, reflecting the progression of human evolution.
In the initial stage, $L_1$, there may not have been a pronounced differentiation between the logical states of $L_1$, $L_2$, and $L_3$. However, as time advanced, humans transitioned from the $L_2$ mode to the $L_3$ mode, driven by a desire to better understand and analyze their experiences. This shift in logic coincided with the rise of human consciousness.
It is important to note that the progression from $L_1$ to $L_3$ does not imply a linear or exclusive process. Instead, it signifies an overall advancement in logical capabilities, incorporating additional elements such as illusions, skills, and detailed analysis. These stages represent milestones in the development of human cognition and the evolution of logical reasoning.

The degree of consciousness experience refers to the level or intensity of awareness and subjective experience that an individual possesses. It represents the extent to which a person is conscious of their surroundings, thoughts, emotions, and sensory perceptions.
The degree of consciousness experience varies among different systems,  for more information, please refer to  Tononi (2004)\cite{25t}, Tononi  and  Christof (2015)\cite{Tononi1} and Lee (2023)\cite{Lee}.

If we classify living organisms into distinct categories, such as humans in one category and cats or ants in others, and denote the set of these categories as $\Sigma$, we can establish a function $f$ that maps each living being in $\Sigma$ to the interval $[0,+\infty)$. In this context, the value of $f(x)$ represents the level of consciousness exhibited by the living being $x$. This level of consciousness encompasses the cognitive abilities and mental experiences of $x$, which contribute to its capacity for perceiving reality and engaging in logical analysis of phenomena.
Consequently, each living being possesses a unique logic that is contingent upon its specific degree of consciousness.
To illustrate this concept, let us consider the examples of a human ($x$) and a cat ($y)$. It is widely acknowledged that humans, in comparison to cats, exhibit a higher level of consciousness when confronted with various phenomena. For instance, humans possess the cognitive ability to comprehend the precise causes of an earthquake through a particular logical framework, whereas a cat lacks this capacity.
Therefore, by virtue of their respective degrees of consciousness, we can infer that $f(y) < f(x)$, signifying that the degree of consciousness for a cat ($y$) is lower than that of a human ($x$). This observation highlights the varying levels of consciousness across different living beings, leading to distinct modes of logic shaped by their individual cognitive capacities.

If we consider the degree of consciousness for cats and humans within the intervals $[0,a)$ and $[0,b)$ respectively, it is evident that $a<b$. Consequently, humans have the ability to analyze the consciousness structure (or pseudo-logical behaviors) of cats to a significantly greater extent compared to cats themselves.
Furthermore, let's assume that conditions $L_2$ and $L_3$ apply to humans with degrees of consciousness within the ranges $[0,c)$ and $[0,d)$ respectively. In this case, it is clear that $c<d$. The primary justification for the superiority of situation $L_3$ over $L_2$ lies in this distinction.
Considering the entire set $\Sigma$ as comprising humans and their ancestors, the precise differences between $L_1$, $L_2$, and $L_3$ become more apparent based on the aforementioned relationships. It is important to note that this process of logical change is a result of the improvement in our sensory capabilities.
As mentioned before, the enhancement of our sensors stems from social life and the development of our skills. It is worth emphasizing that our conscious experience does not undergo instantaneous changes. Therefore, the transition from $L_1$ to $L_2$ and from $L_2$ to $L_3$ occurs gradually over time.
It is important to acknowledge that certain factors, such as incorrect education, illness, or other issues, can impact individuals' mental states and potentially alter some of the aforementioned conclusions.

When a person in the pseudo-logic state $L_2$ interacts or collaborates with a person in the state $L_3$, establishing a common logical language becomes essential for effective analysis of events. The process of reaching a shared logical framework involves aligning and harmonizing their respective perspectives, assumptions, and reasoning methods.
To bridge the gap between $L_2$ and $L_3$, the individuals involved must engage in open and meaningful communication. They need to communicate their thoughts, ideas, and interpretations of events to foster mutual understanding. This communication allows them to identify areas of agreement, as well as areas of divergence, in their logical approaches.
In this process, it is crucial to recognize that the person in state $L_2$ may bring unique insights and perspectives influenced by their inclusion of illusions and acquired skills. Meanwhile, the person in state $L_3$ can contribute their detailed analysis and refined logical process.
By actively listening to each other, acknowledging their differing perspectives, and striving to find common ground, individuals in $L_2$ and $L_3$ can work towards establishing a shared logical language. This language acts as a foundation for evaluating and analyzing events collaboratively, despite their initial differences in logical states.
Ultimately, the goal is to reach a comprehensive understanding that incorporates the strengths and insights of both $L_2$ and $L_3$, leading to more nuanced and well-rounded conclusions.

\section{Corollary}
The structure of human thought and perception is intricately linked to its constituent components, which collectively contribute to a diverse range of perceptions. Our behaviors and our understanding of the environment are significantly influenced by the functioning of these system components.
Consciousness, in my viewpoint, can be seen as the initial term that either enters the system through sensory input or is generated by the system's internal components. This initial translation forms the bedrock of our comprehension of the environment, and the efficacy of the system's components relies on the quality of this initial translation.
The performance of consciousness is contingent upon the capabilities of the system's sensors to receive and transmit information to conscious awareness. By closely examining simpler systems such as plants, insects, or even artificial intelligence, we can glean insights into the workings of more complex systems like the human system.
Through a thorough examination of how systems function, we can approach philosophical inquiries concerning the human mind and our understanding of the world with greater precision. It is essential to acknowledge that notions of absolute reality and absolute truth lack intrinsic meaning. The truth or falsehood of a phenomenon is intricately intertwined with the system that processes it.
In the absence of systems, concepts like correct or incorrect propositions and real or unreal phenomena lose their significance. It is through the existence of systems that these concepts are created and endowed with meaning. Systems provide the framework within which these concepts can be contemplated and understood.
In summary, the structure of human thought and perception relies on its constituent components, which shape a range of perceptions. Consciousness serves as an initial translation that influences our comprehension, and its performance hinges on the capabilities of the system's sensors. By examining simpler systems, we can gain insights into more complex ones. Understanding systems allows us to address philosophical questions with greater precision, recognizing that notions of absolute reality and truth are contingent on the system processing them. Concepts like correctness and reality derive their significance from the existence of systems.

According to the exposition presented in this discourse, two universals emerge as objects of intelligibility for human beings. The first universal pertains to the world that encompasses the breadth of our experiential encounters. It is within this realm that the fabric of our experiences is woven, and it is our sensory apparatus that assumes the role of the creators, serving as the vital interface between this experiential world and the second universal.
The second universal, distinct from the world of direct experience, constitutes the domain in which our experiences reside. It is a realm that transcends the immediate grasp of our senses, yet is accessible through the intricate interplay of cognition. In this interplay, the potential and functionality of our sensory apparatus assume paramount significance, as they serve as the lynchpin connecting these two universals.
Crucially, the potentiality and functionality of our sensory system play a pivotal role in bridging the connection between these two realms. The abilities and limitations of our sensory apparatus determine the information we can receive and transmit, thereby facilitating subsequent cognitive processes. It is through this intricate interplay, facilitated by our sensory system, that we gain an understanding of the universal and its operational mechanisms.
In essence, our comprehension of the universal and its workings hinges upon the potentiality of our cognitive system and the flow of information mediated by our sensory apparatus. These aspects shape our perception and enable us to navigate the interplay between the experiential world and the realm of our experiences, ultimately contributing to our understanding of the intricacies that underlie our existence.

Indeed, the representation of the information received by the system as a set $A$ and the potential of the system as $w$ allows us to conceptualize the system's capacity to perceive and interpret the world as a function of these variables, denoted as $\varphi(A,w)$.
This function, $\varphi(A,w)$, encapsulates the interplay between the information encoded in the set $A$ and the potentiality of the system represented by $w$. It determines the system's range of awareness and understanding of the world, delineating the boundaries within which observations are made and conclusions are drawn. Anything beyond this range remains inaccessible to us, as it falls outside the purview of our perception and cognitive faculties.
It is crucial to recognize that our perception of the world, influenced by the operation of our sensory apparatus, can undergo transformations. These transformations may manifest as the creation or disappearance of perceptual phenomena, such as birth or death, rain or earthquake, and so forth. However, it is important to note that these transformations do not imply a literal creation or disappearance of the underlying experiential universe represented by $\varphi(A,w)$. Rather, they signify alterations or changes in our perception and experience of the world, within the defined range determined by the function $\varphi(A,w)$.
In summary, the function $\varphi(A,w)$ captures the system's capacity to perceive and interpret the world, constrained by the information encoded in $A$ and the potentiality represented by $w$. Transformations in our perception do not imply the literal creation or disappearance of the experiential universe, but rather reflect changes in our subjective experience within the boundaries defined by $\varphi(A,w)$.

Throughout the evolutionary trajectory of human consciousness, distinct stages have emerged, each characterized by its own set of transformative dynamics. In the initial stage, as consciousness began to take shape, humans embarked upon a journey of skill development intertwined with the presence of illusions. Within this phase, individuals often ascribed various phenomena to their own illusions, simultaneously honing their skills and abilities.
However, as consciousness continued to evolve, a subsequent stage unfolded—one marked by the emergence of logical frameworks and systems. These logical systems represented a novel facet of human experience, transcending the limitations posed by illusions. The development of these logical systems was a direct consequence of the expanding repertoire of skills and abilities possessed by humans, enabling them to approach and resolve challenges in a more logical and rational manner.
Crucially, these logical systems, in a remarkable interplay, further augmented the skills and abilities of consciousness. The symbiotic relationship between logical systems and skill development played a pivotal role in propelling human consciousness forward. It is within this intricate dance of logical frameworks and skill enhancement that humans distinguish themselves from other creatures, showcasing a unique cognitive prowess.
In essence, the progression of human consciousness entails an initial stage characterized by the coexistence of illusions and skill development, followed by the subsequent emergence of logical systems. These logical systems, firmly rooted in the continued refinement of skills and abilities, then propel human consciousness to new heights, forging a distinctive demarcation between humans and other beings in the natural world.

In summary, cognition encompasses a complex interplay between material and non-material aspects, influenced by a myriad of material causes. This cognitive process also yields material consequences, both within the system itself and in its surrounding environment.
Our perception of the world is constructed through the intricate capture and processing of images by our sensory organs. These images, once analyzed, can give rise to the creation of new images, forming a continuous chain of interconnected visual representations. Emotions, reasoning, inferences, doubts, and perceptions all manifest as diverse forms of these images, each contributing to the multifaceted tapestry of our cognitive experience. Internal and external projection of these images constitutes the essence of the image creation process.
The significance, rationality, or illusory nature of a phenomenon can be understood as the manifestation of various images that emerge as a result of preceding images. Likewise, the presence or absence of a phenomenon can be comprehended through this same principle of image generation.
The process of image creation is intricately entwined with the influence of external and internal physical and chemical processes within our cognitive system. These processes play a pivotal role in shaping our cognitive experiences and the images we perceive, forming a profound interplay between the material dynamics of our existence and the non-material aspects of cognition.

\vspace{0.2cm}

{\bf Data availability.}  No data were used to support this study.

\vspace{0.2cm}
{\bf Conflicts of Interest.}
The authors declare that they have no conflicts of interest.

\end{document}